# High-Efficiency Silicon-Lithium Niobate Grating Coupler enabled by Fully-Etched Subwavelength Structures

## JING WANG, YANG LIU, SHITONG SU, AND WEIWEN ZOU*

State Key Laboratory of Photonics and Communications, Intelligent Microwave Lightwave Integration Innovation Center, School of Integrated Circuits (School of Information Science and Electronic Engineering), Shanghai Jiao Tong University, 800 Dongchuan Road, Shanghai 200240, China

*Corresponding author: wzou@sjtu.edu.cn

**A new approach of silicon-lithium niobate (Si-LN) grating couplers (GCs) with a fully-etched structure is proposed. The coupling strength is engineered through subwavelength structures, allowing improved directionality. The adjoint-based inverse design method is simultaneously adopted to achieve efficient coupling between photonic circuits and fibers. Considering the birefringence of LN, the proposed Si-LN GCs are optimized and fabricated for TE mode along the crystallographic Z and Y directions. The experimentally measured coupling efficiencies are −3.3 dB and −3.9 dB along the crystallographic Z and Y directions, respectively. The measured 3-dB bandwidths are broader than 54 nm and the back reflections can be suppressed down to −15 dB over 50 nm for both directions. The proposed Si-LN GCs are realized by only single-step Si etching, offering a practical solution for fast device prototyping in Si-LN photonic integrated circuits.**

## 1. INTRODUCTION

Silicon photonics facilitates integrated circuits of compact, scalable and large-volume productions, therefore attracting significant interest in various applications such as optical communications [1,2], optical computing [3,4], and others [5,6]. However, silicon (Si) is a centrally symmetric crystal that lacks a linear electro-optic effect, limiting its capability of realizing large bandwidth modulators. To address this limitation, silicon-lithium niobate (Si-LN) heterogeneous integration techniques have been developed to merge Si photonics with thin-film LN which offers an intrinsic Pockels effect [1, 7-10]. Schemas for Si-LN heterogeneous integration include bonding and deposition methods [8-11], providing attractive strategies for integrating high-performance modulators with mature Si devices.

On the Si-LN heterogeneous integration platform, efficient optical interfaces are required to connect on-chip devices and optical fibers. Edge couplers (ECs) and grating couplers (GCs) have been explored as optical interfaces on this platform [12-18]. While ECs exhibit broadband operation, they require precise alignment and smooth facet preparation. In contrast, GCs extract light from the waveguide and radiate it towards an optical fiber situated over the chip, thus allowing flexible placement on the surface of the chip. In addition, GCs enable relaxed spatial alignment for fiber attachments and support larger numbers of optical ports [19]. Consequently, GCs are preferred for the Si-LN platform.

Recently, several conventional GCs demonstrated on the Si-LN heterogeneous platforms have achieved coupling efficiencies around −4 to −5 dB [15, 16]. Among these demonstrations, Si GCs are typically fabricated using two-step Si etching before thin film LN bonding. To improve the efficiency, a dual-level GC has been proposed, yielding a higher coupling efficiency of −2.78 dB, but at the expense of a three-step Si etching [18]. The multi-step etching in the previous demonstrations not only increase the fabrication complexity, but also introduce possible etching depth deviation and misalignment with the fully-etched waveguide, which can degrade the coupling efficiency. For fast device prototyping, a single-step fully-etched process is highly desirable. However, fully-etched slots feature a strong index contrast in the grating, leading to a significant back reflection and poor coupling efficiency [20, 21].

Taking advantage of the simple single-step fully-etched process, we investigated a subwavelength grating (SWG) approach for the GC design on the Si-LN heterogeneous integration platform. By engineering the material refractive index of subwavelength structures, the diffraction strength of the grating is controlled. We further adopt an adjoint-based inverse design method to optimize nonuniform GCs along different propagation directions, which provides more freedoms to tailor the diffraction field distribution. It results that high coupling efficiencies can be achieved. The simulated peak coupling efficiencies are −1.7 dB and −1.96 dB at C-band for the fundamental TE mode along the crystallographic Z and Y directions, respectively. Our proposed GCs are fabricated by single-step Si etching during Si photonics integrated circuits (PICs) pattering. The fabricated GCs have measured coupling efficiencies of −3.3 dB at 1550 nm and −3.9 dB at 1546.4 nm along the crystallographic Z and Y directions, respectively. The experimental results show 3-dB bandwidths are over 54 nm and Low back reflections are below −15 dB (3%) over 50 nm for both directions.

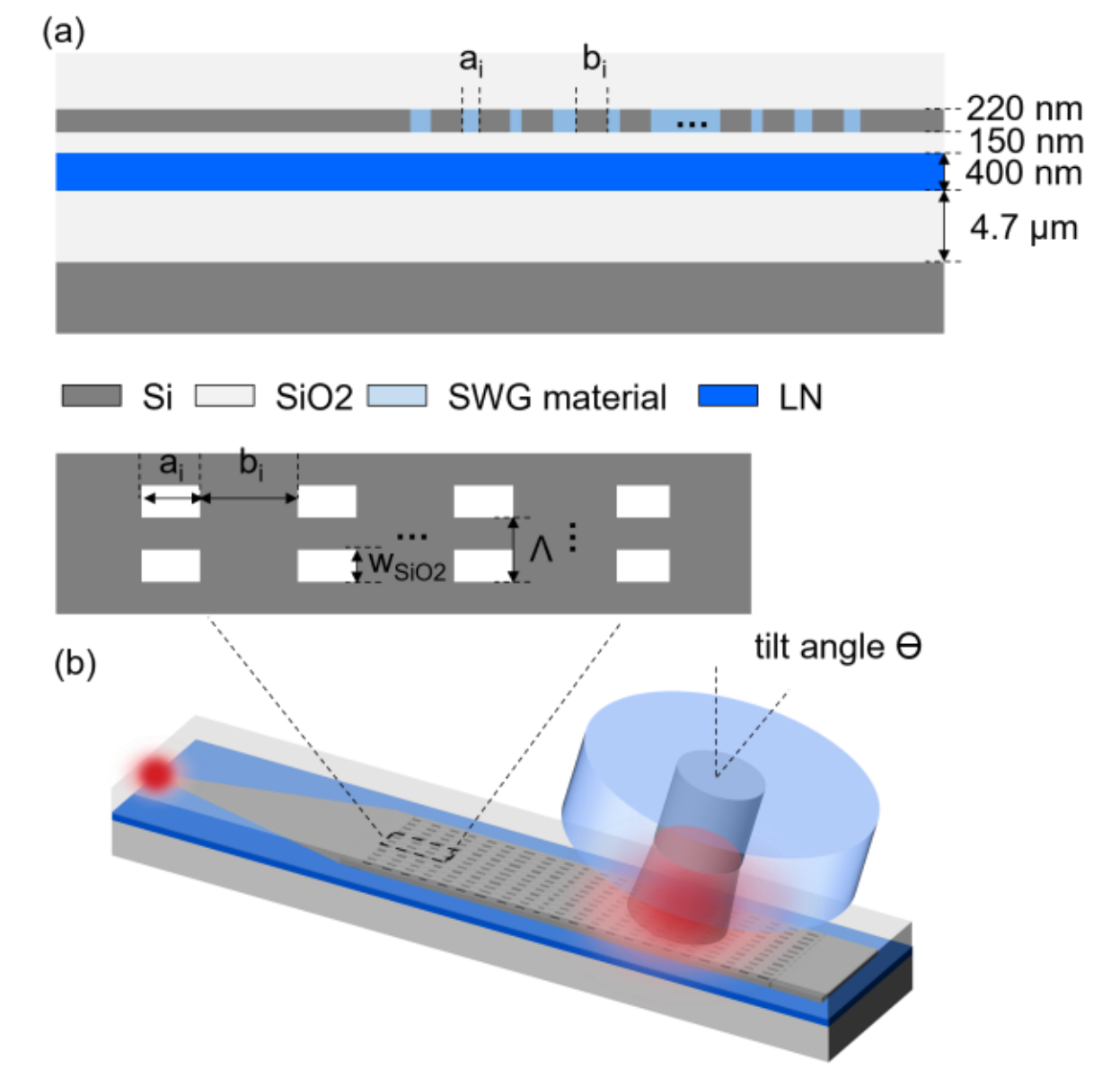

**FIG. 1.** (a) Fig.1. The schematic of the investigated grating couplers on a Si-LNOI (X-cut) monolithic integration platform. (a) cross-sectional and top view with the design details of the GC structures. (b) 3D view with input single-mode fiber (SMF) placed at the angle of Θ.

Figure 1 shows the proposed schematics, including a rectangular GC in the cross-sectional view, top view, and three-dimensional view, respectively. The GC structures are investigated on the Si-LN platform, which consists of a 220-nm-thick Si guiding layer and a 400-nm-thick LN layer (X-cut), separated by an additional thin $SiO_2$ buffer layer of 150 nm. A 4.7-µm-thick buried oxide (BOX) layer is beneath the LN layer, and a 700-nm-thick $SiO_2$ layer serves as the top cladding. Instead of adjusting the etching depth of the two-step etched grating to engineer the coupling strength, we employ the fully-etched subwavelength structures and inverse-designed grating to achieve efficient coupling and simplify the fabrication process.

In this configuration, the grating grooves are filled with the subwavelength structures, formed by the periodic arrangement of silicon segments and rectangular holes, fully-etched in the waveguide core. The rectangular holes are filled with $SiO_2$. By engineering the subwavelength structures, specifically the width of the rectangular hole ($w_{SiO2}$) and the SWG period ($\Lambda_{SWG}$), a wide range of equivalent refractive indices can be achieved. The SWG period is designed to be small enough to suppress diffraction, i.e., $\Lambda_{SWG} < \lambda/2n_{slab}$, where $\lambda$ is the free-space wavelength and $n_{slab}$ is the maximum effective refractive index of the fundamental quasi-transverse-electric (TE) mode of the Si slab waveguide. This allows the structure to act as a homogeneous medium with an equivalent refractive index.

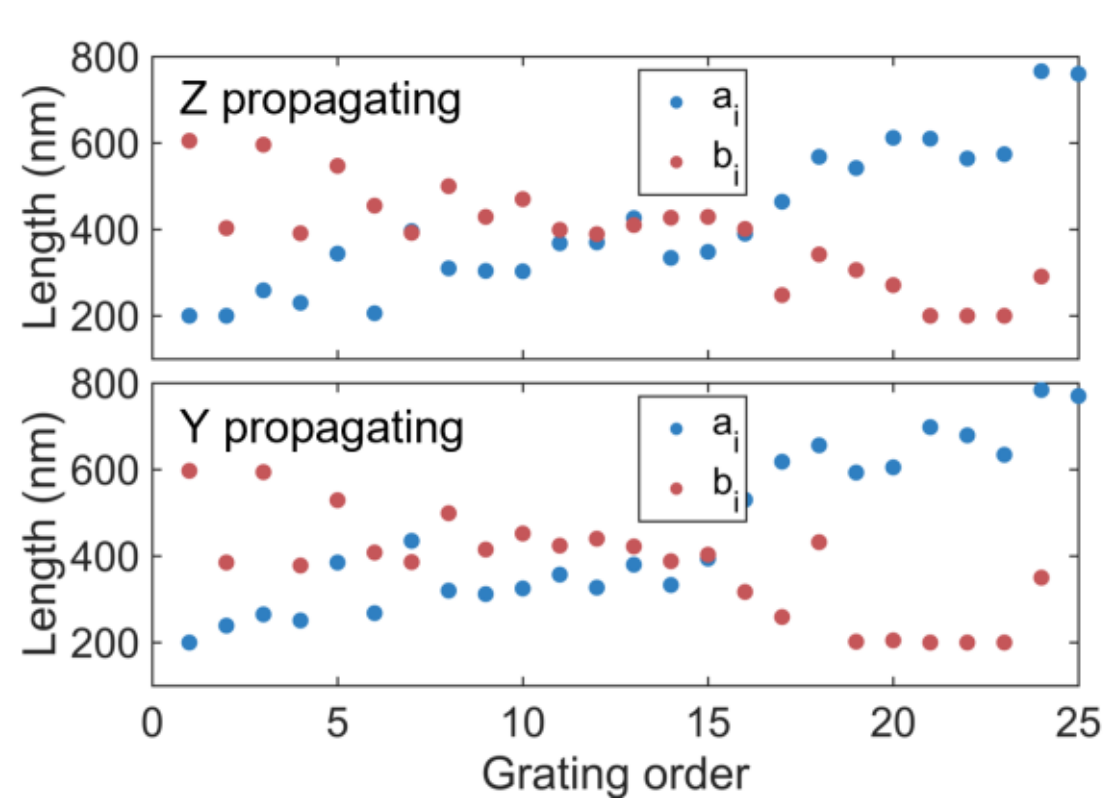

**Fig. 2.** Optimum structural parameters of the Si-LN GCs for Z propagating (upper panel) and Y propagating (lower panel)

The equivalent refractive index, $n_{SWG}$, of the subwavelength structure is determined by the fill factor ($f_{SWG} = w_{SiO2}/\Lambda_{SWG}$) according Effective Medium Theory (EMT) [22]. The formulae for the equivalent refractive index of both the fundamental TE mode and fundamental transverse-magnetic (TM) mode are derived from the second-order EMT method, as presented below [23, 24]:

$$n_{TE}^{(2)} = n_{TE}^{(0)} \times \sqrt{1 + \frac{1}{3}\left(\pi \frac{\Lambda_{SWG}}{\lambda} f_{SWG}(1 - f_{SWG}) \left(n_{SiO2}^2 - n_{si}^2\right) n_{TM}^{(0)} \left(\frac{n_{TE}^{(0)}}{n_{SiO2} n_{si}}\right)^2\right)^2}$$

$$n_{TM}^{(2)} = n_{TM}^{(0)} \times \sqrt{1 + \frac{1}{3}\left(\pi \frac{\Lambda_{SWG}}{\lambda} f_{SWG}(1 - f_{SWG}) \left(\frac{n_{SiO2}^2 - n_{si}^2}{n_{TM}^{(0)}}\right)^2\right)^2} \quad (1)$$

where $\Lambda_{SWG}$ and $f_{SWG}$ are the period length and filling factor of the unit subwavelength structure, respectively. $n_{Si}$ and $n_{SiO2}$ are the refractive indices of silicon and silicon oxide. $n_{TE}^{(0)}$ and $n_{TM}^{(0)}$ are zeroth order approximations of fundamental TE mode and fundamental TM mode, which are calculated by:

$$n_{TE}^{(0)} = \frac{1}{\sqrt{f_{SWG}/n_{SiO2}^2 + (1 - f_{SWG})/n_{si}^2}}$$

$$n_{TM}^{(0)} = f_{SWG} \times n_{SiO2}^2 + (1 - f_{SWG}) \times n_{Si}^2 \quad (2)$$

From the second-order EMT method, the equivalent refractive index is achieved at 2.2 for the fundamental TE mode when $w_{SiO2}$ = 203 nm and $\Lambda_{SWG}$ = 450 nm, which is small enough to suppress the diffraction.

The proposed GC consists of 25 periods, which ensures that most of the light is diffracted vertically instead of being transmitted to the other side of the GCs. Additionally, the thickness of the BOX layer is a critical parameter that influences the directionality of the GC. For a given BOX thickness, careful design of the fiber tilt angle enhances directionality by enabling constructive interference between the upward scattered light from the grating and the light reflected from the underlying dielectric interfaces. In this case, the fiber tilt angle θ is chosen to be 19° based on the BOX thickness of 4.7 µm.

Compared with GCs on the conventional silicon-on-insulator (SOI) platform, it is more challenging to realize a highly efficient GC on the Si-LN heterogeneous platform, since the LN layer underneath the Si layer tends to pull more light downward from the grating region. In order to find the optimum structural parameters of the gratings, the adjoint-based inverse design method is utilized to optimize the lengths of groove ($a_i$) and tooth ($b_i$) of the i-th order grating, aiming for high coupling efficiencies. Full-vectorial two-dimensional finite difference time domain (FDTD) simulations are performed for the Z propagating fundamental TE mode and Y propagating fundamental TE mode, yielding optimum structural parameters for different directions, as summarized in Fig. 2. The minimum feature size is larger than 200 nm, and most of them are above 300 nm. Such gratings enable easy fabrication of the grating even using photolithography, without suffering from practical fabrication impairments.

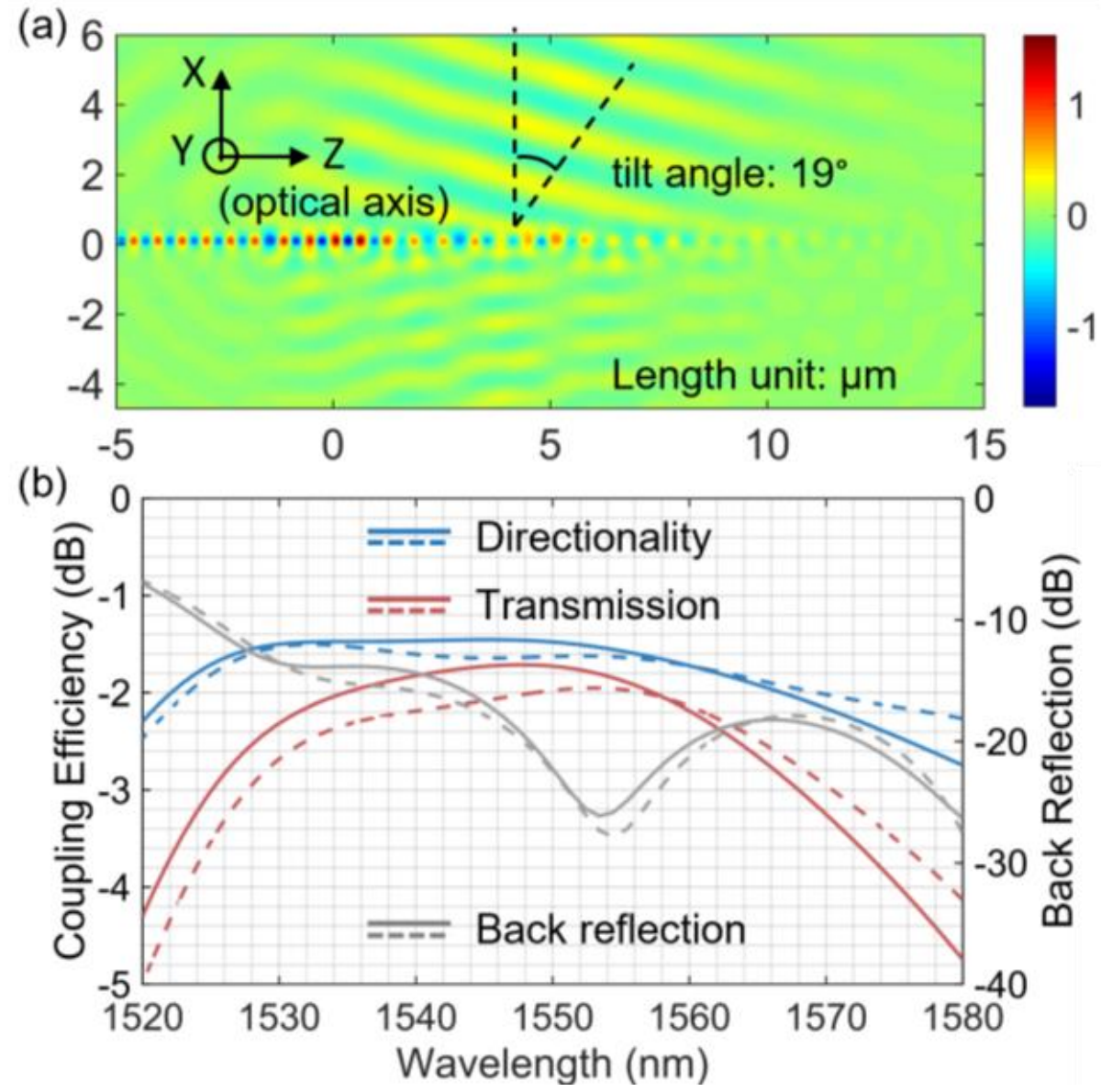

Fig. 3. (a) Electric field component $E_y$ of the Z propagating fundamental TE mode. at the device cross-section, showing most of the light is scattered upward. (b) The simulated spectrum of the optimized GC for the Z propagating fundamental TE mode (solid line) and the Y propagating fundamental TE mode (dashed line).

Figure 3(a) shows the $E_y$ electric field distribution of the Z propagating fundamental TE mode. It can be seen that most of the light is diffracted upward into the fiber when the light enters the grating region. The engineering of individual structural parameters of each grating enables an efficient mode transition and consequently leads to low back reflection. The simulated transmission spectrum of the finally optimized Z propagating Si-LN GC is shown in Fig. 3(b). The maximum upward directionality and the simulated peak coupling efficiency are –1.45 dB and –1.7 dB, respectively. Similarly, the performance of the Y propagating Si-LN GC for the fundamental TE mode is also investigated. As shown by the dashed lines in Fig. 3(b), the maximum upward directionality and the simulated peak coupling efficiency are –1.6 dB and –1.96 dB, respectively. The simulated back reflections are suppressed below –20 dB at the peak wavelengths for both directions.

The test PIC, consisting of pairs of GCs and waveguides, is fabricated on a 400-nm-thick X-cut LN wafer with a 4.7-μm-thick BOX, below which is a silicon substrate. A 150-nm-thick $SiO_2$ layer and a 220-nm-thick amorphous Si layer are sequentially deposited on the LNOI wafer using plasma-enhanced chemical vapor deposition (PECVD) and physical vapor deposition (PVD). Electron beam lithography (EBL) is then utilized to define the pattern of the Si gratings and the interconnecting waveguides. The patterns are subsequently transferred into the amorphous Si layer via inductively coupled plasma reactive-ion etching (ICP-RIE). Here, only single-step etching is required due to the same etching depth for all the on-chip structures. Finally, a 700-nm-thick $SiO_2$ top cladding is grown by PECVD. An optical microscope image of the fabricated GC is shown in Fig. 4(a). The GC is connected to a single-mode straight waveguide with a 250-μm-long taper. The fine structures of the GC are captured by a scanning electron microscope and shown in Figs. 4(b) and 4(c). The Si-LN GC has a width of 15 μm and consists of 20 columns of rectangular holes perpendicular to the propagation direction. Figure 4(c) is the zoom in of Fig. 4(b) for the grating sections, displaying the fabrication details, such as etching width and length.

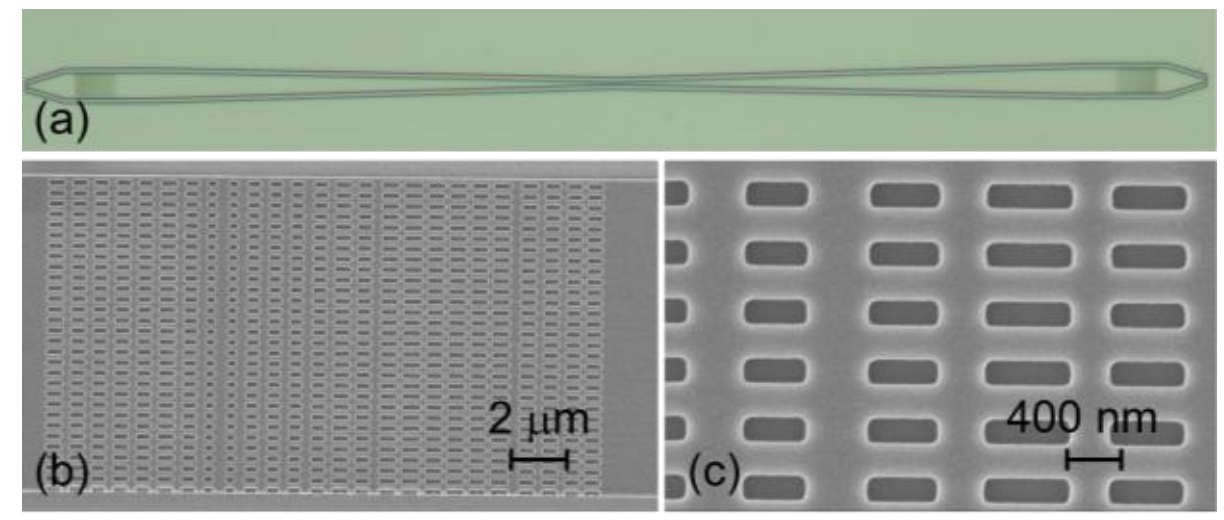

Fig. 4. (a) Optical microscope picture of the fabricated GC. (b) Scanning electron microscope picture of the grating region before the over-cladding growth. (c) The inset depicts the details of the SWG grating.

To characterize the fabricated GCs, the light is coupled into the chip by one Si-LN GC and coupled out by another GC. In this configuration, the light from a broadband laser diode is routed through a fiber polarization controller and then coupled to the GC with a single-mode fiber (SMF). The output light from the GC is collected by another SMF and sent to an optical spectrum analyzer. The input polarization is adjusted to excite the fundamental TE mode in the on-chip waveguide. The fiber-chip coupling efficiencies are determined from the insertion losses measured from the GC pairs, after subtracting the set-up system losses of interconnecting optical fibers. Figure 5 shows the measured spectra for both directions. The measured peak coupling efficiency for the Z propagating GC is –3.3 dB at 1550 nm, with a 3-dB bandwidth over 54 nm. The dotted line in Fig. 5(a) shows that the maximum coupling efficiency and the peak wavelength of the Y propagating GC are –3.9 dB and 1546.4 nm, respectively. The corresponding 3-dB bandwidth also exceeds 54 nm. The measured peak coupling efficiencies are lower than the simulated values, and the measured

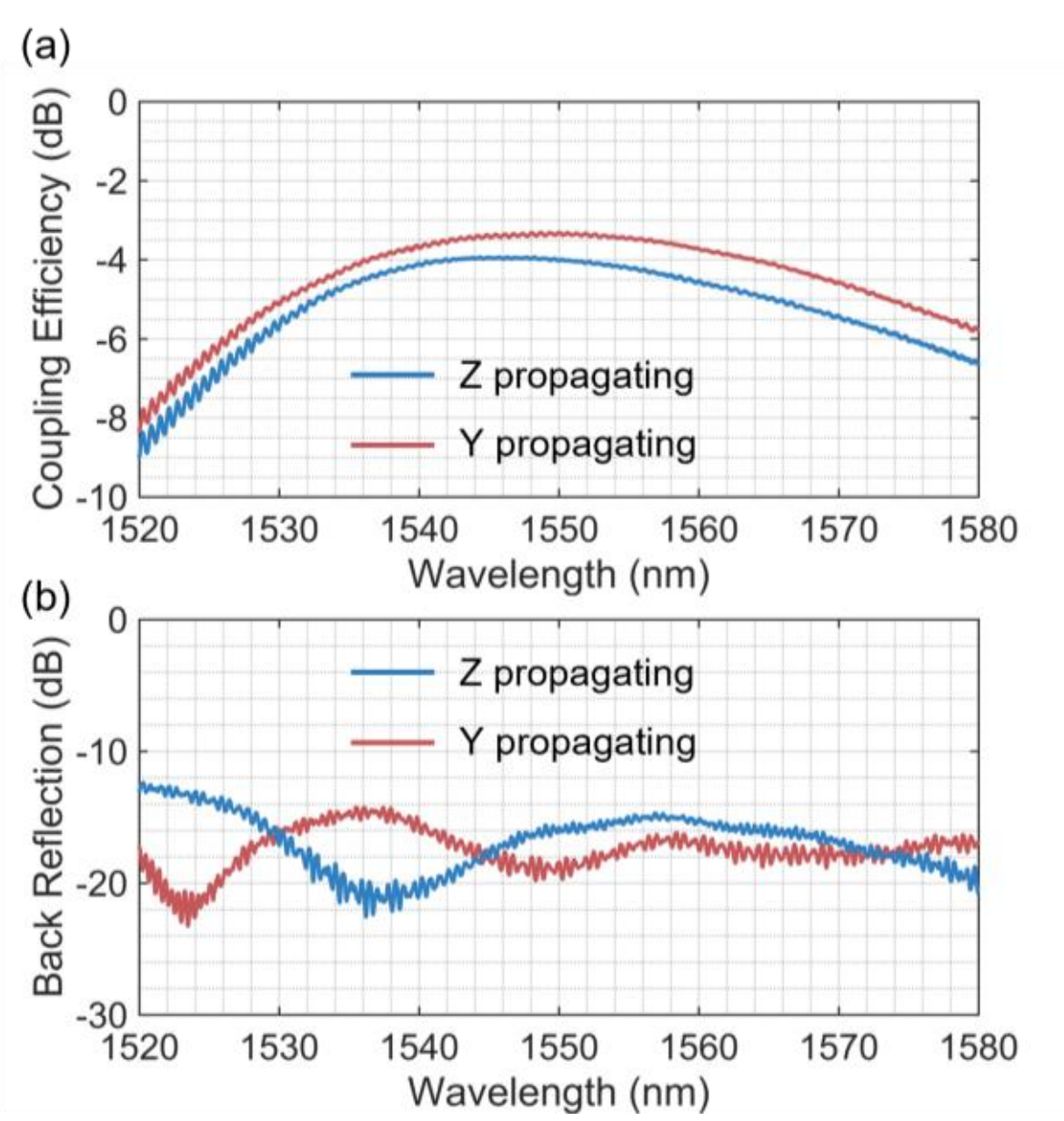

Fig. 5. Measured spectra of (a) coupling efficiency and (b) back reflection along the crystallographic Z and Y directions, respectively.

**TABLE I. Comparison of grating couplers on different heterogeneous LN platforms**

| Platform | Coupler | Process | Efficiency | Ref. |
|---|---|---|---|---|
| Si-LN | EC | SE | –5.2 dB | [12] |
| Si-LN | EC | SE | –4 dB | [14] |
| Si-LN | GC | DE | –4.1 dB | [16] |
| Si-LN | GC | SE | –18 dB | [17] |
| Si-LN | GC | DE | –2.84 dB | [18] |
| SRN-LN | GC | SE | –5.82 dB | [25] |
| SiN-LN | GC | SE | –4.24 dB (Z)<br>–4.02 dB (Y) | [26] |
| Si-LN | GC | SE | –3.3 dB (Z)<br>–3.9 dB (Y) | This work |

SE and DE represent single-step etching and dual-step etching, respectively.

peak wavelengths are also shifted. We think that the observed coupling efficiency penalty and the spectral shift are attributed to fabrication variations and the differences in the parameters, specifically the refractive indices of the different layers. The coupling efficiencies can be further improved by refining the fabrication process and incorporating the measured refractive indices into the design process.

The measured transmission spectra present small ripples with a magnitude range of 0.1-0.5 dB over the wavelength from 1520 nm to 1580 nm for the Z propagating GC and the Y propagating GC, which indicates low back reflections. Experimental results show that the back reflections are below –15 dB (~3%) over 50 nm for both directions. These back reflections are larger than simulated values, probably due to the fabrication variations. Table 1 summarizes the performance of the state-of-the-art optical interfaces on the Si-LN heterogeneous platforms, silicon rich nitride (SRN)-LN heterogeneous platform, and silicon nitride (SiN)-LN heterogeneous platform. It is shown that our proposed GCs represent the first demonstration of Si-LN GCs reaching –3.3 dB coupling efficiency with single-step Si etching process.

In conclusion, we demonstrate the design and experimental characterization of the GCs implanted in the Si-LN heterogeneous platform. The proposed design leverages single-step Si etching, thus easing fabrication. By exploiting the refractive index engineering based on the SWG structures and individual grating structure optimization based on the inverse-design method, we achieve enhanced directionality, high transmissions between fiber and on-chip waveguides, and low back reflections. The proposed GCs exhibit good coupling performance and can be easily fabricated by photolithography with a minimum feature size above 200 nm. Experimentally, the peak coupling efficiencies reach –3.3 dB and –3.9 dB for the crystallographic Z and Y directions, respectively. The 3-dB bandwidths are over 54 nm and the back reflections are suppressed below –3.3 dB over 50 nm for both directions. Benefiting from the simplified fabrication and highly efficient coupling, these GCs are expected to support rapid prototyping for the Si-LN devices and PICs in future high-capacity optical interconnects, microwave photonics, and LIDAR systems.

**Funding.** The National Natural Science Foundation of China (T2225023 and 62405181).

**Disclosures**. The authors declare no conflicts of interest.

**Data availability**. Data underlying the results presented in this paper are not publicly available at this time but may be obtained from the authors upon reasonable request.